\documentclass[usenatbib,usegraphicx]{mn2e}

\usepackage{graphicx}
\usepackage{times}
\usepackage{hyperref}
\usepackage{longtable}
\usepackage{multirow}
\usepackage{natbib}

\begin{document}

\title{Supervised Ensemble Classification of Kepler Variable Stars}

\author[Bass \& Borne]{G. Bass$^{1}$\thanks{E-mail: gbass@gmu.edu}, and K. Borne$^{2}$\\
$^{1}$School of Physics, Astronomy, and Computational Sciences, George Mason University, 4400 University Drive, Fairfax, VA 22030\\
$^{2}$Booz-Allen-Hamilton Inc., VA}
\maketitle

\begin{abstract}
Variable star analysis and classification is an important task in the understanding of stellar features and processes. While historically classifications have been done manually by highly skilled experts, the recent and rapid expansion in the quantity and quality of data has demanded new techniques, most notably automatic classification through supervised machine learning. We present an expansion of existing work on the field by analyzing variable stars in the {\em Kepler} field using an ensemble approach, combining multiple characterization and classification techniques to produce improved classification rates. Classifications for each of the roughly 150,000 stars observed by {\em Kepler} are produced separating the stars into one of 14 variable star classes.
\end{abstract}

\section{Introduction}
\label{sec:introduction}
The study of variable stars has provided a wealth of valuable astrophysical information. Intrinsic sources of variation, such as in pulsation, provide a physical probe and test for our understanding of stellar atmospheres and interiors. These understandings allow for further discoveries, such as the Classical Cepheid period-luminosity relationship \citep{Leavitt_1912}. Extrinsic sources of variation, such as eclipsing binaries or micro-lensing events can also provide valuable information on stellar masses and other attributes, based on the particulars of the variation.

In recent years that has been a rapid expansion in the quality and quantity of data, such as CoRoT\citep{Fridlund_2006}, and Kepler \citep{Gilliland_2010}. Future projects are coming on-line soon, such as the Large Synoptic Survey Telescope (LSST) which will image the entire sky and produce 30 Tb of data nightly \citep{LSST} and Global Astrometric Interferometer for Astrophysics (GAIA), which will catalog around 1 billion stars \citep{GAIA}.

This new era of astrophysical big data offers new advantages and challenges in the important field of variable star analysis. With this expansion of data, the techniques of computational data-mining are essential to maximizing the new resources. Data-Mining is generally divided into two general types. Supervised learning attempts to automatically put new instances  into either existing, already known classifications of those objects. In contrast, unsupervised learning creates new classifications, or looks for correlations in the data irrespective of classification.

\subsection{The Kepler Data}
NASA's {\em Kepler} telescope provides an excellent opportunity to ease into the data-mining paradigm of variable star analysis. {\em Kepler}, during it's primary mission, observed 150,000 stars continuously, measuring the light flux of the star every 15 minutes at unprecedented levels of sensitivity. The observations lasted for a little over four years before mechanical problems ended the original mission \citep{Cowen_2013} although some observations continue to be produced \citep{Howell_2014}. 

The Kepler dataset is broken down by quarters, with each quarter representing 4 months of observations. The one exception is quarter 0, which was preliminary data as the telescope was being tuned. The quarter breakdown is a natural requirement as the telescope is rotated every 4 months to keep the solar panels aligned with the sun. This rotation means that any given star will be observed by a different pixel in different quarters, and thus there are discontinuities in the light curves from quarter to quarter.

Due to time and hardware constraints, analyzing the entirety of the Kepler dataset was not practical. Instead, data was used from Q1-4 and Q16. Using the first four quarters provides a way of looking for patterns with time-scales longer then a single quarter. Q16 was, at the time analysis began, the latest quarter publicly available. It was also analyzed as a way of proving long-term stability of classifications; if our classification of many stars changed after only a few years, that would be indicative of a problem with the classifier.

The details of the mission and instruments can be found in \citet{Koch_2010} and \citet{Caldwell_2010}. \citet{Jenkins_2010} present a discussion of the data-reduction pipeline. All light curves and photometry in this paper is based on the reduced, Single Aperture Photometry (SAP) flux values, publicly available on the MAST database\footnotemark\footnotetext{\url{archive.stsci.edu/kepler/data_search/search.php}}.

The main goal of the mission was to detect earth-like planets by observing the slight dip in light as a planet passes in front of the star. In the process the mission has produced a large amount of data that can be incredibly valuable in a number other areas, including the study of variable stars. 

Presented below is work on verifying and expanding on data mining and variable stars. We present a suite of characterization and classification techniques which together produce an ensemble classifier. Section \ref{sec:Background} presents a discussion of both the state of machine learning and variable stars analysis, including a review on published work combining the two fields. Section \ref{sec:Characterization} discusses time series parameterizations, both in general and the specific choices used in this work. Section \ref{sec:SupLearning} discusses the methodology and supervised learning procedure, while section \ref{sec:SpecificSupLearning} describes each of the supervised learning algorithms used. Section \ref{sec:CombinedAnalysis} presents the combined ensemble classifier and the results produced using that method. Section \ref{sec:Conclusions} presents the work's major contributions to the field, as well as some conclusions, analysis, and thoughts on future work. Finally, section \ref{sec:AddFigs} contains figures showing sample light curves for each of the classes used in supervised learning.

\section{Background and Literature Review}
\label{sec:Background}
As discussed above, data-mining is only just starting to be truly appreciated in astronomy, but it has seen growing usage in recent times. Therefore, it is quite valuable to look into the literature of recent projects that have used these techniques, particularly in the area of classification of variable stars. Below are some of the most salient and interesting examples, discussing methods, results, and lessons that can be taken from the work already done. 

There have been relatively few attempts at automatic classification of variable stars, and only one significant set of attempts using the {\em Kepler} data. This work can be seen in \citet{Blomme_2011} and \citet{Debosscher_2011}, building on earlier work \citep{Debosscher_2007}\citep{Debosscher_2009}.

That work involved training on ground-based data such as TrES Lyr1 (which includes the {\em Kepler} field) light curves, parameterization through a Discrete Fourier analysis, and then construction of a decision tree classifier. They note the importance of classifying known unreliable frequencies, such as the Earth's spin and orbit for ground based telescopes. After this has been taken care of, they select only the significant" frequencies, using the false alarm probability given in \citet{Horne_1986} and \citet{Schwarzenberg_1998}. This contrasts with the authors earlier work in \citet{Blomme_2010} where they took the first 3 independent highest order frequencies. In either case, they then used an automated supervised learning technique to classify the stars into one of 13 classifications. Further discussion of this work is seen in section \ref{sec:TrainingData}. 

Automatic classification of other data surveys have also been attempted. For example, the stars in the Wide-Infrared Survey Explorer (Wise) were classified by \citet{Masci_2014}, using Fourier decomposition to provide attributes for a random forest classifier. 

There have also been several attempts at supervised learning for classifying Hipparcos stars. \citet{Dubath_2011} and later \citet{Rimoldini_2012} present similar methodologies, using both Random Forest and Bayesian Networks to produce classifications. They also discuss using the Random Forest to determine attribute significance, examining over 100 attributes to determine which ones are most helpful for classification. 

Supervised learning has been used in other wavelength domains as well, such as X-ray \citep{Lo_2014}. They use SMOTE, a method of dealing with unbalanced samples to aid in the random forest classification. For more on SMOTE and its use in this work, see section \ref{sec:Smote}. The attributes used include Lomb-Scargle based periodogram analysis as well as more general attributes such as skew and amplitude. 

It is instructive to compare and contrast the different methods utilized by each of these researchers. First, for supervised learning, a training set of "correct" classifications must already pre-exist. \citet{Masci_2014} and \citet{Dubath_2011} use a set of well categorized stars from the Hipparcos catalog, \citet{Rimoldini_2012} used a collection of sources, including General Catalog of Variable Stars \citep{Samus_2009}, the MACHO Variable Star Database \citep{Alcock_2003}, and the All-Sky Automated Survey \citep{Pojmanski_2006}. \citet{Blomme_2010} used a combination of Hipparcos, OGLE, and CoRoT data for training purposes. 

Supervised learning has been used in other wavelength domains as well, such as X-ray \citep{Lo_2014}. While most analysis of variable stars has been limited to supervised learning, \citet{Sarro_2009} present a promising example of unsupervised learning. They perform clustering analysis of the Hipparcos, OGLE, and CoRoT databases. \citet{Stello_2013} perform a classification task on {\em Kepler} data, looking at red giant stars in particular. \citet{Stoev_2013} describe a study of variable stars in the WFCAM Transit Survey (WTS).They used a least squares fitting technique to determine frequencies and up to four harmonics of the previously determined frequencies, identifying several hundred periodic variables. 
\citet{Lopez_2013} discuss the almost 200,000 light curves studied by the ESA mission {\em INTEGRAL}'s Optical Monitoring Camera (OMC).They describe the 18 attributes they use, all but two of which are derived from the light curve (those two are J-H and H-K 2MASS colours). First, they use a fast chi-squared method to determine the periods \citep{Palmer_2009}. \citet{Richards_2012} discusses the use of active learning to improve classification.

Overall, the field of data-mining variable stars is still quite new. Most attempts have confined themselves to picking attributes through Fourier analysis or the related Lomb-Scargle periodograms, and using Random Forest supervised classifiers. Examples of other learning techniques are quite rare, despite their promise.

All of the data analysis discussed below was performed on a personal computer. The five quarters' of data represent approximately 150 GB of data, a large dataset by astronomical means, but relatively small compared with other machine learning tasks. The most intensive process was the characterization (discussed below) which took multiple hours per quarter. Characterization was done in sequence (quarter by quarter), and produced output on the order of 100 MB per quarter. This was manageable for machine learning techniques in reasonable timescales (less than 1 hour for most of the tasks described below).

\section{Data Characterization}
\label{sec:Characterization}
The first step in any attempt at analyzing variable stars is finding some way of parameterizing the light curve. This parameterization involves reducing the light curve into a relatively small set of attributes that describe the data contained within the light curve, essentially a form of dimensionality reduction. The original light curve has a dimensionality equal to the number of observation points. For data like that produced by {\em Kepler}, which is sampled every 15 minutes, this means that even a single quarter of data have a dimensionality in the thousands. Obviously, this high number is very difficulty to deal with and understand meaningfully.

Instead, certain characteristics of the light-curve can be calculated, such as period of repetition, skew, kurtosis, etc. The methods of parameterizing light curves (and more generally, any time series), used in this work are described below. It should be emphasized that these characterizations do not fully describe the light curve, and that some information is lost in the dimensionality reduction. However, a good set of attributes is one in which a maximum amount of usable information is retained while reducing the number of attributes needed to a manageable amount. 

\subsection{Fourier Transform}
\label{sec:FourierVars}
Fourier transforms are a very widely used standard method of describing a function. They transform a function in the time domain into one in the frequency domain by approximating the function as a (potentially) infinite sum of simple trigonometric functions. Each of these trigonometric functions represent the various frequencies that in composite form the original time series. 

Fourier transforms have been used in an incredibly wide range of signal processing problems, including astronomy and astrophysics. In astronomy, the two primary algorithms used for this type of analysis are Least Squares Spectra Analysis (LSSA), also known as the Lomb-Scargle algorithm after the two independent derivations of the algorithm \citep{Scargle_1982}, \citep{Lomb_1976}, and the Discrete Fast Fourier Transform (DFFT) \citep{Cooley_Tukey_1965}. 

While a "true" Fourier transform involves continuous data, in principle most observational data are inherently discrete. Ideally the sampling time is short relative to whatever variability is occurring, but it is incredibly rare to have continuous data. Thus, both of the above algorithms are designed to deal with continuous data. While Lomb-Scargle is considered to be better in dealing with irregularly spaced data, it is considerably more computationally intensive. With over 150,000 light curves to analyze, processing speed was at a premium. In addition, when analyzing only a single quarter, the {\em Kepler} data are typically well spaced, since major gaps only occur between quarters when the telescope rotates. Ultimately the convenience, and more importantly, speed of a DFFT proved more attractive for this work. 

More specifically, the python numpy implementation of the DFFT was used \citep{Numpy_2011}. The Discrete Fast Fourier Transform converts a time series into a set of coefficients of complex sin functions ordered by frequencies. The DFFT is also the algorithm used in the only other attempt at doing classification of all of the Kepler stars \citep{Blomme_2010}, \citep{Blomme_2011}. 

The methodology presented in \citet{Debosscher_2007} was followed closely when doing the Fourier Analysis. Those author later developed an improved methodology, \citep{Blomme_2011}, however the original methodology was described as working fine on the satellite-based data for which it was designed. As this describes the {\em Kepler} data, the original, simpler, methodology was replicated here.

First, linear de-trending was applied to each light-curve to remove telescope systematics that sometimes survived the data-reduction pipeline done by the {\em Kepler} team. Next, a window function was applied to the light curve, to reduce errors at the beginning and edge of the light curve caused by the discrete nature of the function. The Hanning window was chosen as the standard \citep{Testa_2004}.

Unlike \citet{Debosscher_2007}, in which the Lomb-Scargle method was used, we followed instead the example of \citet{Blomme_2011}, in which a Discrete Fourier transform was applied to the light curve, producing a power spectrum. When looking at a single quarter at time, the primary source of data gaps (telescope rotation) is removed, however there were still occasional small data gaps. When present, these were filled using an average of the beginning and end point before the DFFT was applied. The peak finding algorithm from the Python Scipy package \citep{Jones_2001} was used. This method smooths the function with wavelets and then uses those wavelets to identify multiple peaks in order of significance. 

Given the best peak frequency a least-squares algorithm was used to find the amplitudes and phase values that produced the sinusoidal function that had the closest match to the original light curve. This sinusoidal function was then subtracted from the original time series, a process called whitening. The least-squares algorithm was then used again to find a new best frequency, finding a second and then third best frequency. The function was of the form:
\begin{equation}
y(t) = \sum_{j=1}^{4} a_{j,n} \mbox{sin}(2\pi f_n jt)+b_{j,n} \mbox{cos}(2\pi f_n jt) + b_{0,n}
\end{equation}
Where y(t) is the magnitude as a function of time. This produced three equations (one for each frequency), that characterize the light curve. Following the literature, just the first three overtones were included in the fit. Again, it is important to keep in mind that with just the first three of an infinite set, we cannot be said to have perfectly described the light curve. However, it is hoped that we have extracted a significant amount of the information the light curve contains with only a relatively small number of attributes.

These three equations can be combined into a single equation (see \citet{Debosscher_2007} for details). Since this equation is not phase invariant, which is inconvenient for future light curve comparisons, the Fourier components (a$_j$ and b$_j$ from above), are transformed into amplitudes and phases. The amplitudes are inherently phase-invariant, and the phases can be made invariant by arbitrarily choosing ph$_{11}$ as a reference equal to 0. It is noted that these attributes are only phase-invariant for mono-periodic curves. For multi-periodicities, they are not strictly invariant. 

In total, there are three amplitude attributes and four phase attributes, for each frequency, with the exception of the first frequency, which has only three phase attributes, since one of the first frequency's phases is used as the reference and defined to be zero and thus contains no information. In addition, two other attributes are saved.

The first is the ratio of the variances of the residuals (after subtracting the individual frequency sinusoids), and the original variances. The second attribute, not described in \citet{Debosscher_2007}, is the sum of the square of the residuals. These two attributes describe how well the twelve best sinusoids fit the original data.

\citet{Debosscher_2007} did include one additional attribute, the linear slope. In our work, this is not included, as it is likely due to telescopic sources, and is therefore not considered to have any useful predictive power.

In total, there are 28 attribute from this analysis: three frequencies, denoted f1,f2,and f3; 12 amplitudes (a11, a12, a13, a14,a21,etc.), and 11 phases (ph11, ph12...), the ratio of the residuals, labeled as varred, and the sum of the square of the residuals, labeled as res. These 28 attributes were calculated for each of the ~150,000 {\em Kepler} stars, and make up the Fourier parameterization of the data. 

\subsection{Symbolic Aggregate Approximation (SAX)}
\label{sec:SAXVariables}
\citet{Lin_2007} present SAX (Symbolic Aggregate Approximation) a way of symbolically representing time series that allows for dimensionality reduction and indexing with a lower-bounding index. SAX divides a time series into a sequence of "words." Each word is an equal length section of the series, with the letters describing how the curve is increasing or decreasing within that short section. By dividing it up this way, the sequence of words can be used as the parameters.

When using SAX, there are two main user-chosen parameters. The first is alphabet size, and the second is word length. These two parameters divide up the y and x axis of the time series into equal length sections. For a given alphabet size, the time series' intensity/generic y axis is divided into a number of equal sized amounts given by the alphabet size. So an alphabet size of three means that we are looking at those parts of the chunk that are in the lowest third, middle third, or highest third. An alphabet size of 4 would instead divide into quintiles. 

\begin{figure}
\includegraphics[width=0.99 \columnwidth]{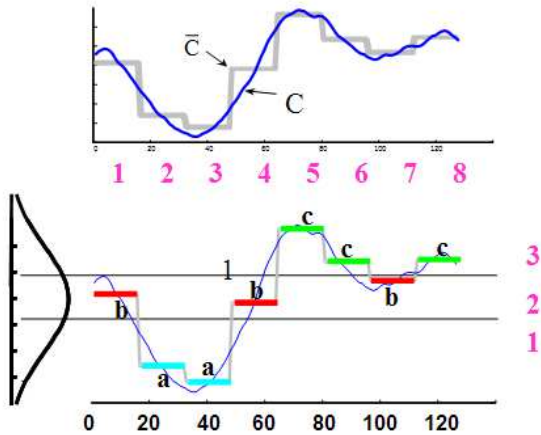}
\caption[SAX Light Curve Description]{Example of how a light curve can be described using SAX. Note that we have made two parameter choices, the word size (top) of 8, and the alphabet size/cardinality of three (below). Figure taken with permission from http://www.cs.ucr.edu/~eamonn/SAX.htm}
\label{fig:SAXExample}
\end{figure}

Similarly, the word length gives the number of equally sized chunks the the 'x' axis (time) is divided into. Combined, the word length and alphabet size give a unique description of a time series. Figure \ref{fig:SAXExample} gives an example of the process.

After running SAX on the dataset, a collection of hundred or even thousands of words (dependent on word length) for each star is produced, a fairly minor reduction in dimensionality in return for a not insignificant loss of information. A longer word length produces fewer resulting dimensions, but more potential loss of signal. The inventors of this algorithm have numerous descriptions of ways to gain useful information from the resulting characterization, a variant of one of them is used here.

A fairly short word length and alphabet size of just three each is chosen, resulting in only 27 possible words, each representing 45 minutes of real time. In addition, since each word is determined independently, some of the potential words are redundant. For example, 'aaa,' 'bbb,' and 'ccc' all represent a flat curve, with the relative height arbitrarily picked. There are several other similar examples. 

The result is just 12 unique words. They are saved as X\#\#\# where the numbers range from 1-3, and are equivalent to 'abc' in the general SAX notation. The possibilities are: X113, X123, X131, X132, X133, X213, X231, X311, X312, X313, X321, and X331. Note that because the relative scale for word is chosen based on just that word, it is also impossible to have a perfectly flat (X111) light curve. Some variance, if only noise, will be detected and represented as a curve.

This is a reasonable number of attributes. For each star a count was made of the number of times each of the 12 possible words appeared. This was then divided by the total number of words to produce a percent frequency count. These 12 attributes thus describe the percent of the time that given word appeared in the star. It was hoped that this information, on its own or in conjunction with the other parameterization might provide useful information, a novel methodology in the field of variable star analysis. 

\subsection{Kepler Photometry}
\label{sec:KepVars}
The other data that were used is information from the Kepler Input catalog \citep{Brown_2011}. Containing astrophysical and photometric data, it was prepared for the purpose of target selection by the Kepler team. Thus, the data in the catalog is included for most of the stars for which there are processed light curves. These data include colour, Temperature, Log G (a measure of the surface gravity of the star), Metallicity, and Radius. We will be using these parameters both to help categorize and characterize the stars, and see if the other parameters can be used to predict the physical parameters.

Because of the large number of stars, no exhaustive spectroscopic survey was done. Instead, the data above comes from broadband and intermediate-band photometry. This was combined with theoretical knowledge, such as atmospheric models. In total, 10 attributes were derived from this data. Their abbreviations, margins of error, and a description can be seen in table \ref{tab:KepPhot}.

\begin{table}
\begin{center}
\caption{Kepler Photometric data. Errors listed from KIC catalog description when available. See \citet{Brown_2011} for further details.}
\begin{tabular}{|c|c|c|}
\hline 
Attribute Name & Error & Description\\
\hline
R.Mag & - & Sloan R Magnitude\\
J.Mag & - & 2MASS J Magnitude\\
KEP.Mag &  0.02 mags & Estimated Kepler Magnitude\\
G.R.colour & - & G-R colour\\
Teff & 200 K & Effective Surface Temperature\\
Log.G & 0.5 dex & Log 10 of the surface gravity\\
Metallicity & 0.5 dex & Log 10 (Fe/H)\\
E.B.V. & 0.1 mags & B-V reddening along line of sight\\
Radius & - & Stellar Radius\\
Total.PM & 20 mas/yr & Total Proper motion\\
\hline
\end{tabular}
\label{tab:KepPhot}
\end{center}
\end{table}

\subsection{Other Attributes}
There was one final set of attributes, using more basic time series and statistical analysis. These attributes were calculate for each star using python numpy packages. Table \ref{tab:MiscVars} lists each attribute and its description. 

The Kepler Photometry and these other attributes are frequently combined into "general" attributes in the later work. These attributes generally describe the stars in different ways than either of the other two sets, and thus are convenient to merge together.

\begin{table}
\begin{center}
\caption{Other attributes from statistical analysis of the light curve magnitudes. }
\begin{tabular}{|c|c|}
\hline
Attribute Name & Description\\
\hline
Std & Standard Deviation of the Flux\\
Median & Median Flux after mean set to \\
& zero during detrending. \\
Amplitude & Amplitude of flux (max-min) \\
skew & Statistical Skewness of the flux distribution\\
kurtosis & Statistical kurtosis (peakedness) of the flux\\
beyond1st & fraction of all data points above the \\
& first standard deviation of flux \\
SSDev & Sum of the Square of the Difference of the \\
& flux from the median flux.\\
MaxSlope & Maximum difference in flux between 2 points\\
Median Slope & Median value of difference between \\
& every pair of adjacent points\\
AbsMaxSlope & Absolute value of MaxSlope\\
\hline
\end{tabular}
\label{tab:MiscVars}
\end{center}
\end{table}

\section{General Supervised Learning Details}
\label{sec:SupLearning}
\subsection{Training Data}
\label{sec:TrainingData}
Supervised learning involves the placement of new data into already 
known classes. In addition to requiring the existence of these classes classifications, there must also be a training set of sample objects with previously made class assignments. There are a few possible approaches that could have been made to producing this training set. 

The first is to manually classify a reasonable subset of the stars in the field, or find in the literature manually classified stars and use this subset as a training set. The second is to train based on pre-classified stars that are not in the Kepler field, but have similar light curve data from having been observed by similar missions such as Hipparcos and OGLE. The third is to use classification data from another study that has already attempted automatic classification of the entire field.

This work primarily uses the third method for training data, using the only existing set of variable star classifications for the entirety of the Kepler dataset, the work of \citet{Blomme_2011}. As a cross validation, the first method is also described, using a comparison some classifications of the {\em Kepler} stars found in the literature.

\subsubsection{Blomme et al.}
\citet{Blomme_2011} build on previous work \citep{Blomme_2010}, \citep{Debosscher_2011} in developing and applying a supervised learning algorithm based on multi-variate Bayesian statistics to produce a multi-stage classification tree. 

Their training set includes light curves from Hipparcos, OGLE, and CoRoT. The light curves are parameterized by searching for significant frequencies and overtones during Fourier Analysis. After performing a DFT (Discrete Fourier Transform) they find up to two significant frequencies and three harmonics, first checking to see if these are "significant" and reliable. The exact definitions of significant and reliable are discussed in their work, and are related to signal-to-noise ratios and common instrumental false frequencies.

Using this method, they were able to provide classifications of every star in the Kepler field into one of the sixteen categories given by \citet{Blomme_2011} in table \ref{tab:BlommeNum}. These classifications will be used as the training data for all of the supervised learning presented here.

One key step in any machine learning is developing a method of analyzing the validity and correctness of any learning results. In this case, because the training set is itself based on the results from a different supervised learning algorithm, any problems in that data set are likely to be replicated in any algorithm discussed below. This is a significant problem, and one of the major incentives for the decision to investigate unsupervised methods as well.

\citet{Blomme_2011} list both their best classification and the classifier's percent certainty in the answer. To try and reduce the weight of erroneous classifications, all stars whose classification was considered less than 90\% certain was removed from consideration. 

This does mean that the classifier produced is not learning on the "difficult cases." This may increase error on those cases. Thus, the results should be considered more a proof of concept and a tool for quickly and fairly accurately classifying most stars, rather than a perfect tool that can completely replace human expertise.

In addition, when presenting analysis on the fraction of cases correctly classified, the correct classification is based on \citet{Blomme_2011}'s best classification. It is theoretically possible that the classifications presented below could have accuracies either better or worse than the ones given had they been compared with manual classifications. While it may seem difficult to imagine how the classifications could be improved, it is possible the combination of multiple learning algorithms picked up patterns from the correctly classified stars and then successfully applied them to incorrectly classified stars.  

\subsubsection{Other Classifications}
To measure this effect, an examination of the literature was conducted for other classifications of the Kepler stars. While there were no exhaustive classification attempts like \citet{Blomme_2011}, various groups have classified sub-sections of the Kepler field, usually searching for specific classes of variable star. \citet{McNamara_2012} examine 252 B stars, sorting 100 of them into $\beta$ Cephei, Slow Pulsating B stars (SPBs), and Binary/Rotation classes. \citet{Slawson_2011} presents a list of several thousand Eclipsing Binaries.  \citet{Uytterhoeven_2011} analyzed 750 Kepler A-F stars based on parameters likely to be $\delta$ Scutti and $\gamma$ Dor class stars, but manually classified all of the stars in the sample, including into other variability classes. \citet{Tkachenko_2013} identify a few dozen $\gamma$ Dor as well as some $\gamma$Dor-$\delta$ Scutti hybrids as well. The hybrids are not used for this validation, but the pure $\gamma$ Dor stars are. 

These classifications were merged, and classes not in the Blomme classification were removed, to provide a more direct comparison. This provided classifications of 2091 stars that could be directly compared with the results of \citet{Blomme_2011}, and the results presented here. Table \ref{tab:ManClassNumbers} shows the number of stars by classification (using the manual classification) in the studied described above. 

\begin{table}
\begin{center}
\caption[Manual Classification Frequency Counts]{Frequency Count of Manual Classifications of Stars.The class abbreviations are defined in table \ref{tab:ManClassNumbers}}
\begin{tabular}{|c|c|}
\hline
Classification & Number\\
\hline
BCEP & 3\\
DSCUT & 162\\
ECL & 1762\\
GDOR & 67 \\
MISC & 11\\
ROT & 56\\
SPB & 29\\
\hline
\end{tabular}
\label{tab:ManClassNumbers}
\end{center}
\end{table}

Because this sample is based on available manual classifications, it is not statistically representative of the sample in general. The majority of the stars are Eclipsing Binaries, and several categories of stars are almost completely absent, most notably Misc/non variable stars, which is the largest in the actual data. Because of this, these classifications are unsuitable for any training purposes, and validation results are somewhat questionable. However their independence from other classifications make the validation worthwhile, even with these weaknesses.

\begin{figure}
\includegraphics[width=0.99 \columnwidth]{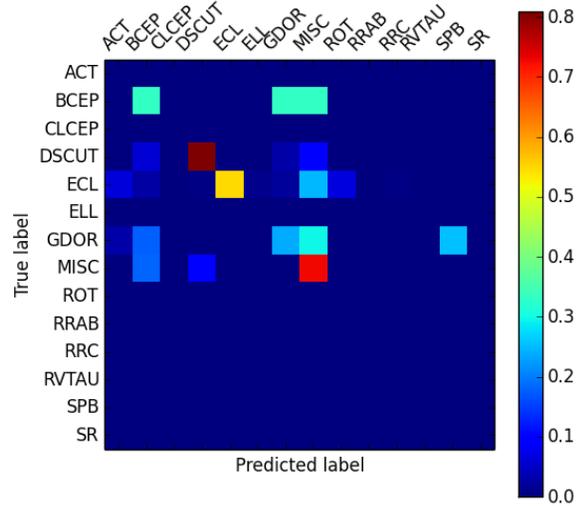}
\caption[Blomme vs. Manual Classifications Confusion Matrix]{Scaled Confusion Matrix for \citet{Blomme_2011} automatic classifications and manual classifications from various literature sources.  Here the manual classifications are labeled as "True Label" and the Blomme as the "predicted", unlike in all other such figures were the Blomme classifications are the "True Label." The colour displays the percent of all objects with a given true classification that were assigned the listed predicted classification.}
\label{fig:BlommeOther}
\end{figure}

Comparing these results with \citet{Blomme_2011} is not particularly favorable. Figure \ref{fig:BlommeOther} shows the confusion matrix between the automatic and manual classifications. The Blomme data "correctly" predicts the classification just 55\% of the time. While \citet{Blomme_2011} do list probabilities for their classifications, the 2091 star comparison sample lists a median probability of the given classification of 0.997, and a mean probability of 0.949. While manual classification has weaknesses as well, in light of the relative poor match between the manual classification and the automatic classifications, some doubt must be placed on the automatic classifications and their certainty.

In comparison to these results, the classifications by the combined method, discussed in this paper in the following sections, increases the "correct" classification to 65\%, a 10 percentage point increase. 

\subsection{Unbalanced Data and SMOTE}
\label{sec:Smote}
When performing unsupervised learning on variable stars, one major issue is unbalanced data. When classifying the variability of a large set of stars, such as the Kepler sample, around 60\% of them are classified as Misc. These stars are either non-variable, or don't fit into any known variability class. While these stars may be intrinsically interesting, particularly in looking for previously unknown classes, they present another problem as well, namely the problem of unbalanced data.

Essentially, even the most common classes (such as ACTs and $\beta$ Cephei) are only approximately one out of every 7 of the stars, far dwarfed individually by the number of Miscellaneous stars, which on its own is more than half of the entire sample. See the \citet{Blomme_2011} classification frequency counts, shown in table \ref{tab:BlommeNum}. This means that for supervised learning, there is a very strong bias towards guessing that a star is Miscellaneous. This problem can be even more acute for inducing the classifier into avoiding coming up with the rarer classifications, such as the various types of RR Lyrae. 

There are two main solutions in the machine learning literature to this problem. The first involves assigning a cost for misclassifying a rare class as a more common class. When the algorithm attempts to minimize error, a misclassification of, e.g. a RRC as a MISC might count as being as much as a miss of 5, 10 or even hundreds of misses in the other direction. The main problem with this approach in the context of this research is our desire to use multiple different learning algorithms on the same sample and compare them. Applying this misclassification error would need to be done individually for each method, requiring a substantial amount of effort and would risk the cross validity of the different methods.

Fortunately, the second main solution is more applicable. This involves re-sampling the training data set, either by oversampling the minority class, under-sampling the majority class, or both. Both involve intentionally statistically biasing the training set, either by creating duplicates of the less common types, or by removing large numbers of the majority class. 

Either way, the resulting data contain more equal numbers of each class, at the expense of this added bias. Under-sampling removes valuable data, and in the case of some of our rarer classes (where there are only a few total samples) there will either still be persistent differences in number of classes, or there will be so little data machine learning techniques will be useless. Oversampling meanwhile runs the risk of encouraging over-fitting as multiple copies of the exact same object are created.

One of the most commonly used solution to these problems is a hybrid method called SMOTE \citep{Chawla_2002}. SMOTE, or Synthetic Minority Over Sampling Technique over-samples the minority class, but rather than creating exact copies, instead creates new, synthetic minority class objects. These objects are created by taking an example of a minority class, and then looking at that object's five nearest neighbors. A new synthetic member of the minority class is then chosen by picking attributes within the range of the six object set  of the original minority class plus the five nearest neighbors.

For example, to double the sample size a random 1 of the five nearest neighbors are chosen. Then a random point in the n-dimensional feature space (where n is the number of attributes describing the point) along the line connecting the original object and the neighbor is chosen as a new synthetic object with the same class as the original object.

Using this method, we found a marked improvement in the classification, particularly of the minority classes. Section \ref{sec:RFSmote} discusses the resulting improvement in classification using SMOTE with the Random Forest classifier, subsequently all other classifiers were used exclusively with synthetic SMOTE training data. While the method is not perfect, and does result in many stars classified as MISC by \citet{Blomme_2011} instead being classified as variable stars, primarily ACT and BCEP, the overall improvement seems worth the potential downsides.  

It is also not completely certain that the "mis-classifications" are in fact erroneous. We may be finding more variable stars than \citet{Blomme_2010} was able to identify. Furthermore, it is generally better to falsely identify non-variable stars as variables than to miss some stars that actually are variable, something that using SMOTE will tend to encourage. In any case, even if these classifications are erroneous, it does seem that the advantages outweigh the disadvantages.

\begin{table}
\begin{center}
\caption[Blomme et al. (2011) Classes and Frequency Counts]{Number of Kepler stars with given classification by Blomme et al. (2011). Note the extreme number of Miscellaneous compared with other classes, and some classes (such as SR, RVTAU, etc.) with only a few examples in the entire dataset. Section \ref{sec:AddFigs} Shows example light curves for each of these classes.}
\begin{tabular}{|c|c|c|}
\hline
Abbreviation & Class & Number \\
\hline
ACT & Activity & 18250 \\ 
BCEP & $\beta$ Cephei & 18245 \\ 
CLCEP & Classical Cepheids & 18 \\ 
DSCUT & $\delta $ Scuti & 884 \\ 
ECL & Eclipsing Variables & 2322 \\ 
ELL & Ellipsoidal & 191 \\ 
GDOR & $\gamma$ Doradus & 424 \\ 
MISC & Miscellaneous/non-variable & 90388\\ 
ROT& Rotating & 7718 \\ 
RRAB & RR Lyrae type AB & 4 \\ 
RRC & RR Lyrae type C & 16 \\ 
RVTAU & RV Tauri & 3 \\ 
SPB & Slowly Pulsating B Star & 371 \\ 
SR & Semi-Regular & 4 \\ 
\hline
\end{tabular}
\label{tab:BlommeNum}
\end{center}
\end{table}

\subsection{Over-fitting}
One significant problem in machine learning is the issue of over-fitting. Over-fitting occurs when the algorithm learns to identify the noise, instead of the signal, in a particular problem. This can arise when there are too many free parameters, or too many different classification techniques are applied to the same data-set. This is particularly problematic if the sample set itself is small, as random parameters coinciding with categories can swamp any hidden signal.

One way to reduce over-fitting, is to rigorously separate the training and fitted data. This was done for each of the below methods, where a subset, typically 80\% of the data were removed and only the remaining 20\% used for training. The resulting classifier was then run on the removed data, reducing the opportunity of over-fitting falsely inflating the classifier's precision.

\subsection{Attribute Use}
One area of interest was the relative value of the different light-curve characterization attributes. Although there were a total of 61 attributes, they fell into three main groups. The first were the Fourier phase, amplitude attributes and frequency, which uniquely describe the 3 most significant Fourier frequencies, as described in section \ref{sec:FourierVars}. The second is the SAX attributes (described in section \ref{sec:SAXVariables}), and the final are the misc, which include both attributes found via Kepler photometry (section \ref{sec:KepVars}), and the other attributes describing the curves variability statistically, such as skew and kurtosis.

For each learning algorithm, four classifications were produced. The first was with all of the 61 attributes, and then the remaining three were with just one of the above sets of attributes. This allows comparison not just between the classifiers, but also between the characterization attributes. 

\section{Methods and Results}
\label{sec:SpecificSupLearning}
\subsection{Random Forest}
In order to understand the Random Forest algorithm, one must first understand the Decision Tree that it is based on. In a decision tree algorithm, the machine learning algorithm designs a series of tests that slowly narrow down the object to be classified. One advantage of this algorithm is that it produces results that are simple to understand and interpret. Multiple classes can be compared for "similarity" based on how close they are to each other on the tree. They are a basic standard of this type of analysis, and have been used in several of the studies mentioned above (\citet{Blomme_2011}, among others.

Considered by many to be the gold standard of machine learning, Random Forest algorithms are one of the most commonly used methods of machine learning, both for variable star classification, and in general in the field of machine learning. A Random Forest randomly creates a large number of decision trees (a group of trees being a forest), and then the "best" one is selected, based on the percentage of correct classifications. However, in the process of creating many unique decision trees, further information is gained. For example, the trees can be compared to see which attributes have the most predicative power. However, for data where there are multiple levels of classification (for example, double-mode Cepheids are a type of Cepheid which are themselves a type of ellipsoidal variables), there can be a bias towards those variables with multiple levels.
\subsubsection{Dimensionality Reduction and Attribute Significance}
\label{sec:VarSig}
One of the advantages of the Random Forest algorithm is its natural ability to produce information about the importance of each of the attributes for the final classification. Because each decision tree in the forest uses a different, randomly chosen subset of the attributes, if the forest is sufficiently dense, then for each attribute an analysis can be made of the performance of trees with or without just that attribute. 

While this could be done artificially in any classification scheme by simply removing that particular attribute, in the random forest, we also see the attribute present or absent in the context of different groups of other attributes. This may be significant in cases where, for example, a attribute is highly significant when paired with a different attribute, but not useful on its own. 

The Random Forest algorithms used, in Python and R, have built in methods for determining attribute significance. The two main methods are the mean decrease in the Gini Coefficient \citep{Gini_1912}, and the mean decrease in accuracy. 

One problem in machine learning is the so-called "Curse of Dimensionality." This refers to the problem that when the number of attributes describing each data point becomes large, distance functions start behaving differently than they do in low-dimensional regimes. The exact definition of a "high dimension" is algorithm specific, but can be as low as double digits of attributes. Since we have 60 attributes describing each star, it is thus important to consider whether some algorithms may be adversely affected by the curse of dimensionality.

The curse can mostly clearly be looking at how a Euclidean distance function changes at higher dimensions. The distance function is used as the basic method of most simple clustering techniques, as a way of determining if two objects are similar (and thus likely to be in the same class) or different (and thus likely to be in different classes). The Euclidean volume V of a hyper-sphere of radius R and dimension D is given by:
\begin{equation}
V = \frac{2 R^D \pi^{D/2}}{D\Gamma(D/2)}
\end{equation}
A hypercube of similar dimensions and length would have Volume:
\begin{equation}
V = (2R)^D
\end{equation}
Thus, if you imagine inscribing the hypersphere into the hypercube, and comparing their relative volumes, you get:
\begin{equation}
\frac{V_{sphere}}{V_{cube}} = \frac{\pi^{D/2}}{D2^{D-1}\Gamma(D/2)}
\end{equation}
Which approaches zero as D increases without bounds. In other words, at high dimensions, the hypercube's volume increases relative to the inscribed hypersphere. This means that for any given object, most other objects will appear in the corners of the space, and all neighbors are equally distant. This makes algorithms such as the k-nearest neighbors understandably problematic. 

The Gini Coefficient is a measure of the variance in a frequency distribution. In the context of a Random Forest, it represents the increase in homogeneity of the child nodes relative to the parent nodes. That is, if the parent node had an equal number of objects of Class A and class B (perfectly heterogeneous, with a Gini index of 1), and the node perfectly separated the classes into two child nodes one that was fully Class A and the other that was fully class B (fully homogeneous, with a Gini index of 0), the decrease in the Gini Coefficient would be 1. 

For any given tree, the mean decrease in Gini Coefficient for a specific attribute is found by taking the decrease in Gini coefficient for every branch in the decision tree that involves that attribute. The net change is summed over the tree, and then normalized. For the entire random forest, the net change in Gini Coefficient is again summed and normalized, producing a single attribute, where the greater the mean decrease, the more significant the attribute.

By permuting a specific attribute over all trees and determining the mean decrease in accuracy/Gini coefficient, one can find a measure of the importance of that attribute. If the attribute was an important part of the classification, then by randomizing it, the correct classification rate should decrease, as all information contained in that attribute is lost. The final score is produced by taking the mean error before and after the permutation over the whole forest, and then normalized by the standard deviation of the errors.

While the Random Forest algorithm is providing crucial information about the attribute significance, an examination of the literature does throw up some warning signs. \citet{Strobl_2007} shows that statistical bias shows up in attribute importance when different types of attributes (i.e. categorical vs. continuous) are used, or categorical attributes with different numbers of categories. Fortunately, our attributes are all continuous, so this is unlikely to be a problem.

More worrisome, \citet{Archer_2008} and \citet{Nicodemus_2010} show that their can be a preference towards attributes that are correlated with one another. While some of our attributes may be correlated, its is unclear how significant of a bias this is. Ultimately, even if the attribute significance is not perfect, imperfect knowledge seems better than none.

Thus, if we can find just the 10 or so most significant attributes of the 60 that we have used for classification, there is an excellent chance this will significantly improve clustering and unsupervised learning later on. As the random forest has proved to be one of the consistently best classification methods (see section \ref{sec:CombinedAnalysis}) and has built-in attribute significance testing methods, it was ideal for this purpose.

For each quarter of data, the R Random Forest built in attribute significance function was calculated. Both Mean Decrease in Gini Coefficient and Mean Decrease in Accuracy were found, and added together, producing a single attribute describing the significance of each attribute (previous tests had shown that adding the two together produced similar results to looking at either one separately). The median significance attribute for each quarter (Q1,Q2,Q3,Q4, and Q16)was then found, producing one totaled significance per attribute for the entire data set. The results are seen in table \ref{tab:VarSig}

It is interesting to see that each of the three main sets of attributes have fairly high importance. The Fourier components f1,f2,and f3 are highest, with a11 and varred being up there as well, vindicating the work of others that this method of analysis is quite useful. However, the SAX attribute X123 is also in the top 10, and several others are in the top 20. Other attributes, such as statistical ones like kurtosis and fbeyond1std and Kepler photometry attributes like Teff and Radius are also useful. 

In total, this seems to vindicate the approach of combining many different characterizations. One wonders whether other characterizations might also have useful information and further improve classification results if they are only attempted.

\begin{table}
\begin{center}
\caption{Shows the relative significance of the 20 most significant attributes, as well as the standard deviation of that significance.}
\begin{tabular}{|c|c|c|}
\hline
Attribute Name & Attribute Significance & Std. Dev \\
\hline
f2 & 6.258 & 1.020 \\
f3 & 5.649 & 2.828 \\
f1 & 4.049 & 1.623 \\
 kurtosis & 2.511 & 2.595 \\
 a11 & 2.391 & 1.309 \\
 varred & 2.027 & 2.018 \\
 fbeyond1std & 1.521 & 1.849 \\
 X123 & 1.049 & 1.388 \\
 SSDev & 0.983 & 0.693 \\
 std & 0.934 & 0.621 \\
 Teff & 0.748 & 0.904 \\
 res & 0.627 & 0.450 \\
 median & 0.307 & 0.551 \\
 Radius & 0.297 & 0.565 \\
 R.Mag & 0.191 & 0.495 \\
 G.R.colour & 0.153 & 0.805 \\
 KEP.Mag & 0.145 & 0.598 \\
 Log.G & 0.076 & 0.501 \\
 J.Mag & 0.060 & 0.623 \\
 skew & 0.053 & 0.663 \\
\hline
\end{tabular}
\label{tab:VarSig}
\end{center}
\end{table}

\subsubsection{Random Forest and SMOTE}
\label{sec:RFSmote}
\begin{figure}
\includegraphics[width=0.99 \columnwidth]{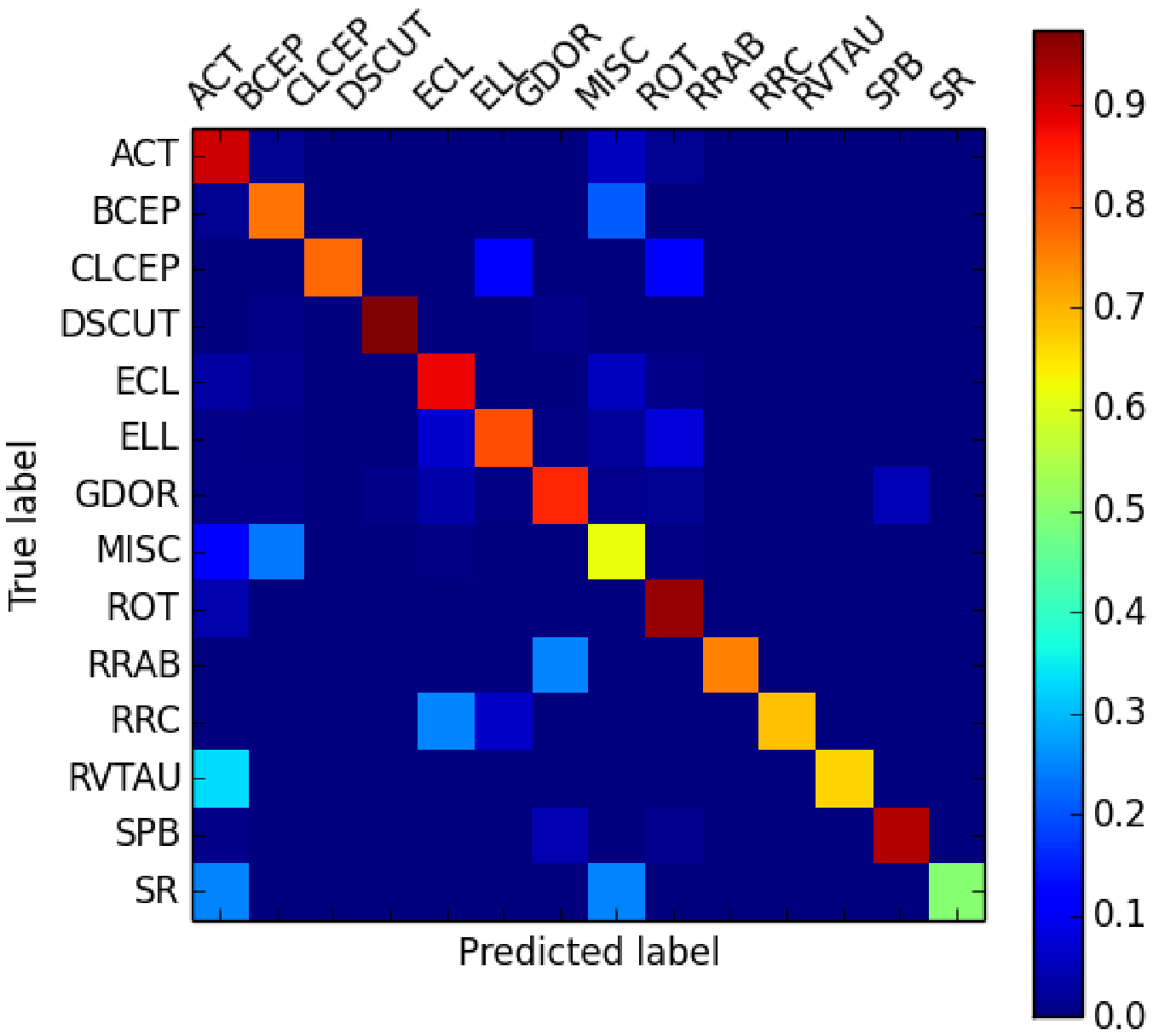}
\caption[SMOTE Random Forest Confusion Matrix]{Heat map of Scaled confusion matrix for classifications with SMOTE on the training data. The colour displays the percent of all objects with a given true classification that were assigned the listed predicted classification.}
\label{fig:SmoteComp}
\end{figure}

\begin{figure}
\includegraphics[width=0.99 \columnwidth]{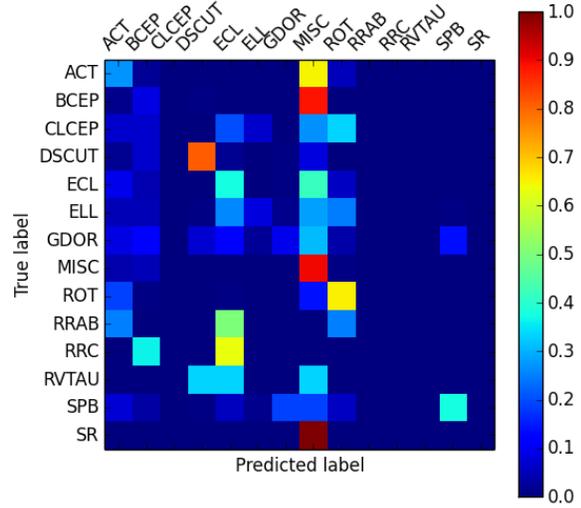}
\caption[Random Forest Confusion Matrix Without SMOTE]{Heat map of Scaled Confusion Matrix for classifications without SMOTE on the training data. The colour displays the percent of all objects with a given true classification that were assigned the listed predicted classification.}
\label{fig:SmoteCompNoSmote}
\end{figure}

As discussed in section \ref{sec:Smote}, synthetic data were produced to provide a training set that was less biased towards the most common classification. As a comparison, a random forest was also run on the original data. As can be seen in figures \ref{fig:SmoteComp} and \ref{fig:SmoteCompNoSmote}, the SMOTE results provide a significant improvement on classifications, particularly the less common types. 

Note that without SMOTE, most of the errors were stars being incorrectly classified as MISC. With SMOTE, the classification generally improves, albeit with some increased error in MISC stars being classified as various other classes, primarily ACT and BCEP. All of the attributes were used in this analysis.

In particular, the classification using all attribute without SMOTE produced a correct classification rate of just 68\%. In addition, that is strongly weighted towards correctly classifying Miscellaneous stars at the price of mis-classifying non-Miscellaneous stars. In contrast, using all of the attributes with SMOTE on the same stars and with all attributes produced a correct classification rate of 71\%. The errors are also tilted more towards false negatives of MISC/false postives of other classes, which is more useful for future work. Given this improvement and the strong theoretical arguments in favor of its use, SMOTE was used for all further supervised learning work. 

For the above comparison, there were actually three types of training data used. The first case didn't use SMOTE at all. Instead, roughly 20\% of the data were used to train, and then that produced a random forest classifier that was applied to the remaining 80\% of the data. Different divisions of training vs testing were applied, each producing roughly similar results to what is presented here.

The second trial used SMOTE as the source of training data and the actual data for testing. In addition, while SMOTE helped even the amounts of each of the types, the rarest types were still significantly less represented in the training sample than the more common types. However, all of the most common classes were equally common. The exact numbers of objects (simulated or real) used for training after applying SMOTE are shown in table\ref{tab:SmoteNumbers}

\begin{table}
\begin{center}
\caption[SMOTE Frequency Counts]{Number of simulated (via SMOTE) stars of each category in the training set. Using SMOTE produced more even numbers of each classification, allowing better classification of rarer classes and avoiding over-classifying stars as Misc.}
\begin{tabular}{|c|c|}
\hline ACT & 19874 \\ 
\hline BCEP & 19956 \\ 
\hline CLCEP & 19992 \\ 
\hline DSCUT & 1137 \\ 
\hline ECL & 18430 \\ 
\hline ELL & 19881 \\ 
\hline GDOR & 19712 \\ 
\hline MISC &  20000\\ 
\hline  ROT& 17694 \\ 
\hline RRAB & 8001 \\ 
\hline RRC & 19998 \\ 
\hline RVTAU & 3928 \\ 
\hline SPB & 19840 \\ 
\hline SR & 3991 \\
\hline 
\end{tabular}
\label{tab:SmoteNumbers}
\end{center}
\end{table}

Finally, a different use of SMOTE was used to produce a third training set. This set was produced from only 80\% of the original, and consists of simulated data only for all of the classes except Misc (because the MISC category was an outright majority, SMOTE was not needed to avoid any over-fitting problems). This meant that we could test on the remaining 20\% with no chance of over-fitting and exactly equal weights for each class in the training data set. This third method was the one used in all of the subsequent analysis, for both the Random Forest classifiers as well as the other methods.

\subsubsection{Random Forest Results}
\begin{figure}
\includegraphics[width=0.99 \columnwidth]{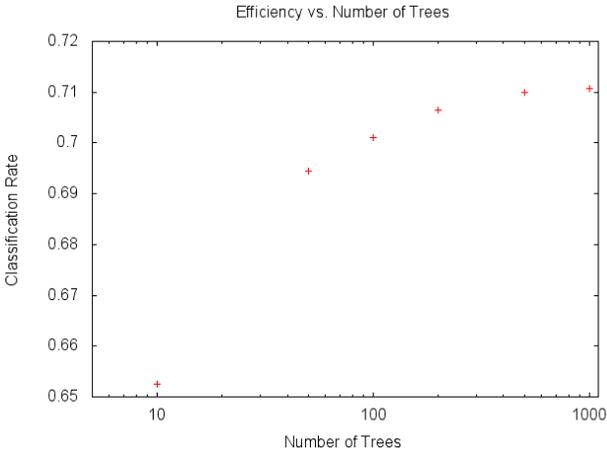}
\caption[Random Forest Number of Trees vs. Classification Rate]{A plot of the correct classification rate (as measured by comparing with the classification by Blomme) vs. the number of trees used in the Random Forest. Random forest is trained on the SMOTE data, and tested on Q1 data. Note that improvement has almost completely leveled off after 500 classifiers, despite steadily increasing computational time.} 
\label{fig:NumTreesEff}
\end{figure}

After producing a training data set using SMOTE on the quarter 1 data, a Random Forest was produced. After multiple trials, a forest size of 500 was found to produce the best combination of accuracy and speed. Figure \ref{fig:NumTreesEff} shows the classification improvement as the number of classifiers increases. It is clear that there is little improvement as the number of classifiers increases beyond 500, justifying our forest size choice. Computational time was on the order of hours when trying to classify more than 500 trees.

Overall the Random Forest results, particularly when using all of the attributes, produced very good results. When classifying based on Quarter 1 testing data, the "correct" (Blomme) classification was produced a substantial 71\% of the time. Considering the likely uncertainties in the test data themselves, this is quite impressive. 

\begin{figure}
\includegraphics[width=0.99 \columnwidth]{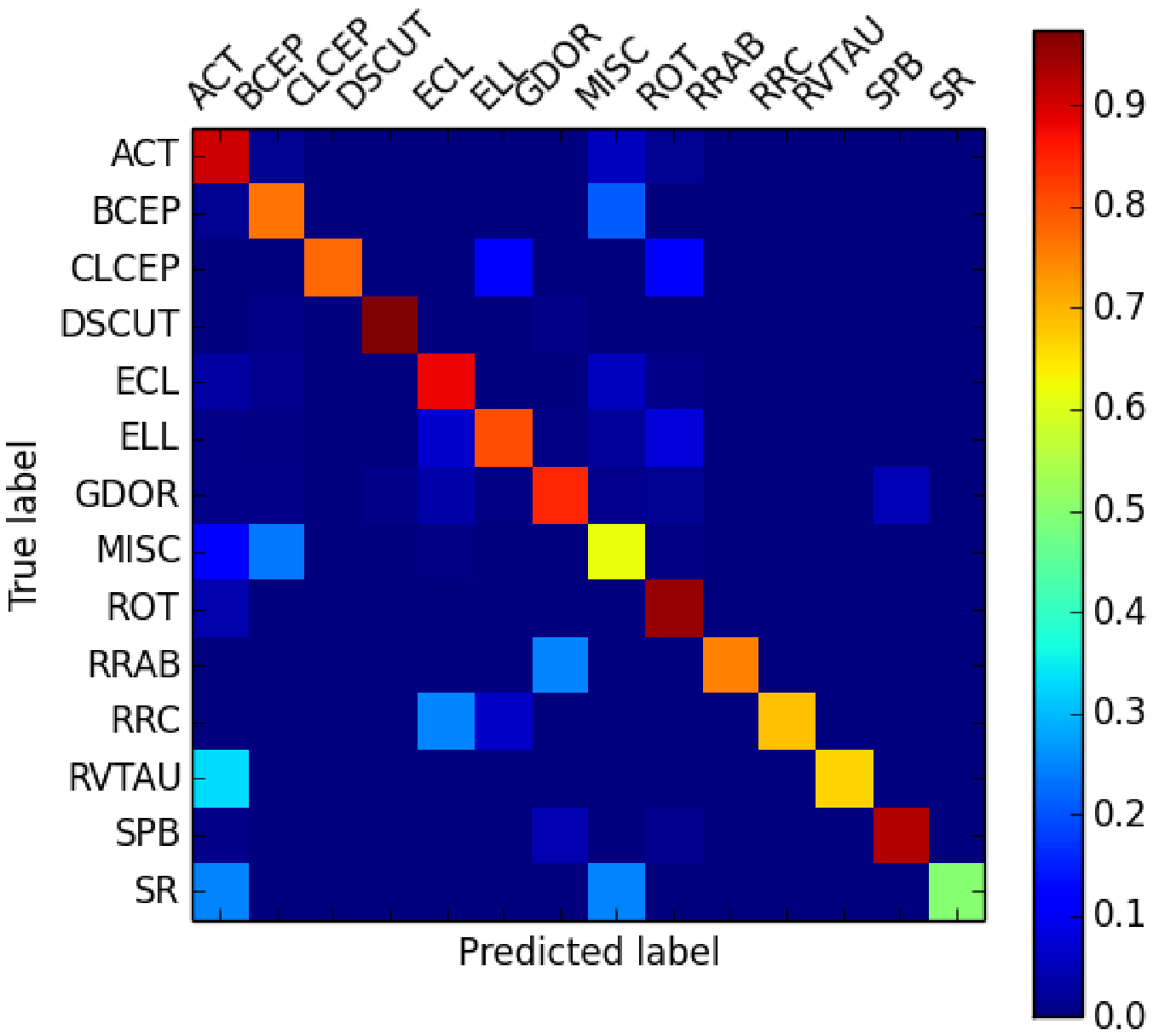}
\caption[Random Forest Confusion Matrix]{Heat map of Scaled Confusion Matrix for Random Forest using all attributes, trained on SMOTE and tested on Q1. There are clear indications of diagonality, as well as some significant errors, mostly involving both false positives and false negatives for the Miscellaneous category. The colour displays the percent of all objects with a given true classification that were assigned the listed predicted classification.}
\label{fig:Q1RFConfusionMatrix}
\end{figure}

The classification with just the Fourier attributes was 68\%, with just the SAX attributes was 62\%, and with the miscellaneous attributes was 68\%. The confusion matrix with all of the attributes is show in figure \ref{fig:Q1RFConfusionMatrix}

The classification rate drops considerably for later quarters. For example, there is only a 49\% match using all of the attributes with the Quarter 2 data as test data. Quarters 3, 4, and 16 produce similar values.

\subsection{Supervised Bayesian Network}
\label{sec:SupBayesNetwork}
Bayesian Networks are a graphical data exploration device that provide probabilistic relationships between attributes based on Bayes theorem. The name was coined by Judea Pearl, who was instrumental in developing and describing their properties and capabilities \citep{Pearl_1988}. One of their primary advantages is their simplicity and easy interpret-ability based on their visual nature; relationships between attributes are easily spotted with a visual inspection.

Bayesian Networks provide a powerful tool for discovering relationships between attributes, using Bayes theorem. At a basic level, a Bayesian Network is composed of points, one for each attribute describing the data-space, and arcs that represent relationships between those attributes. For example with stellar parameter data, Mass and Temperature would each be points, and would likely have an arc between them, in accordance with the main sequence relationship.

For this application of Bayesian Networks, the first step is binning the 61 parameters. The Bayesian Network paradigm is designed for discrete data, and cannot work with continuous data like the original attributes. Instead, the data have been separated into five bins, representing very low, low, medium, high, and very high for each of the attributes. Each bin is equally sized, taking 1/5 of the total Kepler stars, thus the size of the bins for each parameter depends on that attribute's relative range.

Each attribute now represents a unique graph point, and the Bayesian Network calculates correlations between the various characterizations based on their relative frequencies in the Kepler data. Bayesian Networks can provide many different learning results, including both supervised and unsupervised. The results here focus mainly on supervised learning, but offer some interesting possibilities in unsupervised learning. 

To perform supervised learning, the Bayesian Network is trained on a subset of training data, and then optimized based on its predictions of one particular attribute, namely Class (V1). Implementation of the Bayesian Network was done using GeNIe (Graphical Network Interface), \footnotemark{\footnotetext{https://dslpitt.org/genie/})  a software package that creates Bayesian networks with an assortment of options and graphical interfaces. For all of the work with GeNIe, SMOTE was used to prepare training data as described in previous sections, and the data were again binned into five equally sized bins. 

Because Bayesian Networks are not necessarily supervised learning, there are options to create the network to minimize overall error or the error of a specific attribute. For supervised learning, picking a specific parameter to optimize is the correct choice. Thus, the search algorithm was chosen to optimize the classification based on the correct prediction of the 'Class' parameter, labeled V1. 

One of the adventages of the Bayesian network is that it simultaneously finds inter-attribute relationships. In this paradigm, first the network structure is built, optimizing for correct classification of the V1 class attribute, but allowing for inter-relationships between the other attribute. 

There are a few free parameters for the Bayesian search. The key ones are Max Parent Count, Number of Iterations, and Number of Test Data-Folds. The Max Parent Count is fairly self-explanatory, limiting the maximum number of parents any given node (attribute) can have. The limit is primarily for computational reasons. Next, the Number of Iterations is the number of different times the whole process is restarted, to avoid finding only local minimas. Finally, the last significant parameter is the number of folds of independent training data that were used. This helps further minimize any risk of over-fitting.

There are also two parameters involving probability of links, and the Link Probability and Prior Link Probability. These were left unchanged from the defaults of 0.1, and 0.001 respectively.

Multiple trials showed that a max parent count of 8-10 was usually sufficient, and beyond that would slow down the algorithm considerably for insignificant improvements in classification accuracy. Similarly, a limit of 500 iterations fell into the sweet spot of computational time and accuracy. Together with 3-4 folds, these were the main parameters used in training the Bayesian Network structure.

Early results, with lower numbers of iterations and parents produced very poor results, but eventually they leveled out. The final results improved significantly, compared with both previous results and the naive Bayes search results. The final true positive classification rate using all of the attributes was 66\% (91687/138838), with most of the error being due to over-classification of stars as Misc.  

Equally interesting, this method also produced some very interesting visualizations of the attributes inter-relationships. Figure \ref{fig:BNAllViz} shows the Bayesian Network produced from training on Quarter 1 SMOTE data. The groupings of parameters as discussed in section \ref{sec:Characterization} are clearly distinguishable in this network. 

\begin{figure*}
\includegraphics[width=2 \columnwidth]{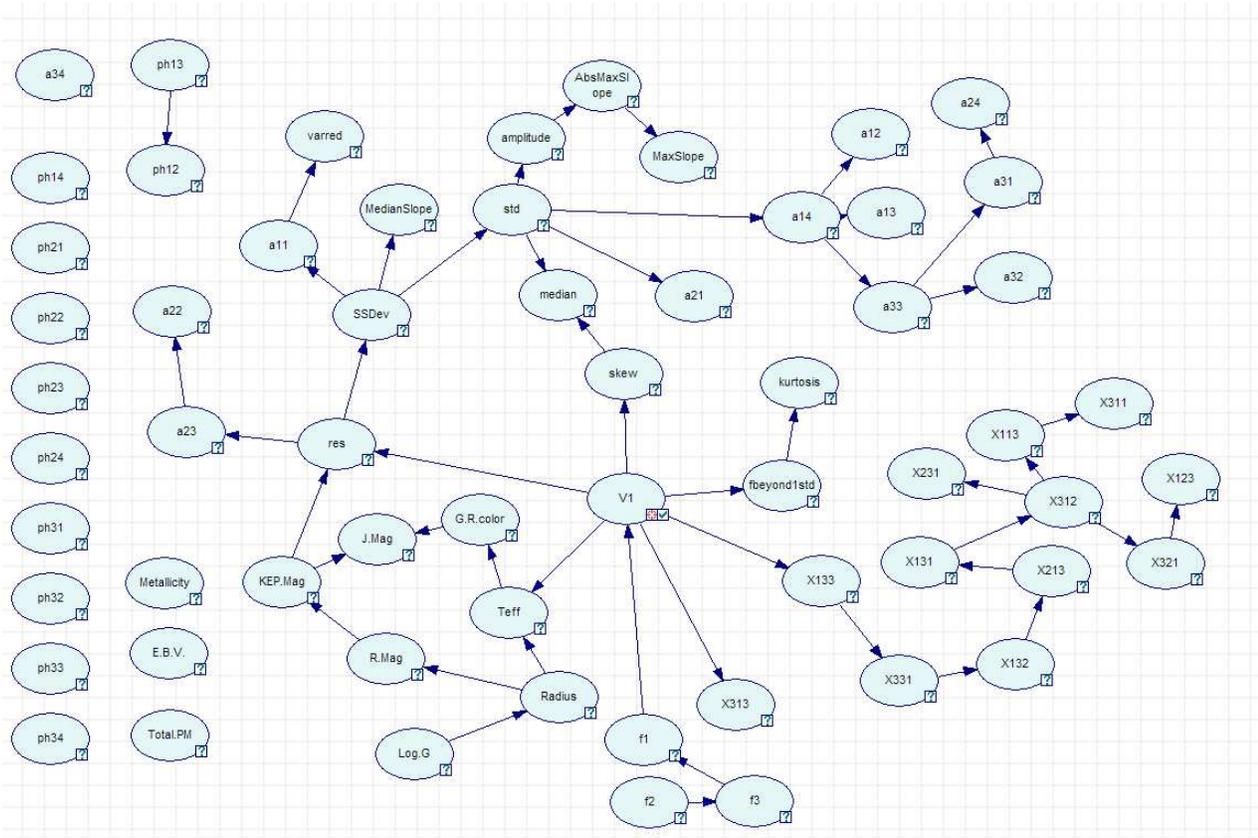}
\caption[Advanced Bayesian Network Visualization]{Bayesian Network produced using a Bayesian Classifier trained on SMOTE Q1 Kepler data. The network is trained to optimize classification of V1, the class attribute. The direction of the arrows is arbitrary, simply indicating a Bayesian correlation between the two attributes.}
\label{fig:BNAllViz}
\end{figure*}

The bottom right of the figure shows almost all of the SAX attributes have strong relationships with one another, and through that, with the classification attribute V1. In the bottom left, we see relationships between the photometric magnitudes, colours, temperature and radius. That these attributes are tightly linked is not surprising, but is a good confirmation that the network is finding physically realistic information.  

The network also shows that the Fourier phase attributes (ph) are generally unimportant, not strongly linked to each other or towards any other attributes, something that also showed up in the Random Forest attribute significance work. 

The top of the diagram shows some interesting correlations between some of the time series descriptive attributes, such as skew, and maximum slope and the Fourier amplitude attributes, while the most important Fourier attributes, the three most significant frequencies are all correlated with each other and through f1, with the class attribute.

Overall, examining the visualization of the network offers some common sense assurance that there is physical reality being captured by the model. It also gives some tantalizing hints at unexpected relations, and also serves to provide some confirmation for the significance testing done in section \ref{sec:VarSig}, when discussing Random Forests.

\subsection{Neural Network}
Neural Networks are a family of machine learning methods inspired by human cognition. A series of individual nodes, analogous to human neurons, are interconnected and allowed to send signals from one to the next. Some form of input is provided, which then propagates through the network based on the network's geometry of connections, before finally outputting to a specific exit node.

In the case of this work, the input is the 61 attributes describing a star. This set of data is passed to the neural network, which after some processing, produces a single output node that assigns the star one of the known variable star classifications. As this is supervised learning, the geometry of the neural network is developed based on training data, as well as the specific neural network algorithm being used.

\subsubsection{Multi-Layer Perceptron}
The Neural network used in this work is the Weka implementation of a Multi-layer Perceptron (MLP). In an MLP, there are three or more layers of nodes, an input layer, an output layer, and one or more hidden processing layers. In the processing layer(s), each node receives signals from one or more of the nodes in the layers behind it, and sends signals to one or more of the nodes in front of it. 

Each node has its own, non-linear activation function. This function performs a weighted addition of the signal from all of the previous layer's nodes that are connected to a given node, and then converts it to an output signal, typically between 0 and 1. That output signal is then sent to all of the connected nodes in the next layer. 

The weights and node connections are what is learned by the algorithm, typically using the back-propagation method. When running the algorithm, these details are less immediately apparent than the general parameters inputted into the program. The primary parameters include training time (in epochs), learning rate, momentum, and a binary flag indicating whether or not to use a decaying learning rate. 

After multiple experiments, the default classification rate of 0.3 without decay was found to be the best for this problem. These tests were trained on the SMOTE data, and then tested on quarter 1. A slightly lower momentum, of 0.1 (the default is 0.2), seemed to produce slightly better results, and thus was used exclusively for all further experiments. 

In addition, there are two options to end learning before the pre-set number of epochs. The first involves a Graphical User Interface (GUI) and allows the user to watch the classification and manually stop when it appears to have plateaued. The second method involves withholding some percentage (25\% was used) of the data as a validation set. After each stage, the classification rate of the validation set is checked. When this rate stops improving, the program ends.

While the GUI was used experimentally a few times, the validation set method was used to produce the final results. This method introduces less human-bias, and was found to produce consistently equivalent classification results while requiring less human intervention.

\subsubsection{Neural Network Results}
First, a neural network was trained on the Q1 SMOTE data. A training time of 25 epochs was allowed, along with a 25\% validation percentage. The validation option being turned on means that the algorithm checks regularly for improving classification, and the training can end either at the limited number of epochs (200), or when the classification ceases to improve, whichever occurs first. 

Overall, the results were very consistent, regardless of which set of attributes were used for classification, producing the correct classification somewhere between 20-50\% of the time, but usually 30-40\%. The best single classification was 51\%, for quarter 3 using just the Fourier attributes. 

The Fourier attributes produced the best classifications on average, slightly edging out using just the general set of photometry and statistical attributes. The general set was, however, more consistent, classifying between 27-38\% correct, while the Fourier attributes classified anywhere from 25-51\%.   The worst, as usual was the SAX attributes, with only a 16\% correct classification rate. 

\section{Ensemble Classifier Results}
\label{sec:CombinedAnalysis}
After providing classifications for each method separately, one combined ensemble classification was produced by taking the most common classification. It is well known in the literature \citep{Polikar_2006} \citep{Rokach_2010} that ensemble classifiers combining many individual classifiers, even if the classifiers are individually poor, often produce excellent results. 

Each of the above classifiers was trained, on the Q1 SMOTE data, using all of the attributes, just the SAX attributes, just the Fourier attributes, and just the Misc. attributes (all attributes that were not SAX or Fourier). However, the Random Forest, Bayesian Network, and Neural Network each do, producing a total of 12 classifications for each star. Then, the classification most commonly picked was chosen as the ensemble classification.

In addition to enhancing the classification results, this method has the additional benefit of giving some measure of the certainty of the classification. A star that is classified by a majority of the classifiers the same is more likely being classified correctly; thus if one is looking for specific types of stars, they can rank potential targets for manual follow-up by likelihood. This is described further below, in section \ref{sec:AoS}

\subsection{Overall Classification Rates}
Not surprisingly, since the classifiers were trained only on the Q1 data, they were most successful in that classification, matching Blomme's classification 76\% of the time. This percentage also is significantly better than any of the individual classifiers. 

Quarter 3 provided the next best classifications, significantly better than Quarters 2, 4, and 16, which all had approximately the same rate. The reason for this is a bit unclear, although it is clear that most of the improvement is from the Neural Network classifier. Table \ref{tab:SupByQuarter} shows the classification rate by quarter and classifier.

\begin{table}
\begin{center}
\caption[Classification Rate by Quarter and Classifier]{Classification rate by quarter and classifier. The Classifier acronyms stand for Random Forest, Bayes Network, and Multi-Layer Perceptron (Neural Network), respectively. Also included are the attributes used, General, Fourier, SAX, as described in sections \ref{sec:KepVars}, \ref{sec:FourierVars},\ref{sec:SAXVariables} respectively. 'All' means that all 61 attributes were used.}
\begin{tabular}{|c|ccccc|}
\hline
 & Q1 & Q2 & Q3 & Q4 & Q16\\
\hline
RFGenRate & 0.682 & 0.482 & 0.431 & 0.446 & 0.468\\
RFFourRate & 0.684 & 0.196 & 0.381 & 0.192 & 0.213\\
RFSaxRate & 0.621 & 0.0248 & 0.060 & 0.0275 & 0.023\\
RFAllRate & 0.710 & 0.448 & 0.545 & 0.469 & 0.550\\
BNGenRate & 0.236 & 0.233 & 0.199 & 0.588 & 0.666\\
BNFourRate & 0.587 & 0.418 & 0.455 & 0.416 & 0.448\\
BNSaxRate & 0.623 & 0.628 & 0.620 & 0.626 & 0.640\\
BNAllRate & 0.669 & 0.651 & 0.652 & 0.651 & 0.629\\
MLPGenRate & 0.378 & 0.305 & 0.335 & 0.298 & 0.271\\
MLPFourRate & 0.426 & 0.257 & 0.508 & 0.246 & 0.251\\
MLPSaxRate & 0.331 & 0.056 & 0.328 & 0.056 & 0.037\\
MLPAllRate & 0.389 & 0.204 & 0.394 & 0.197 & 0.141\\
CombinedRate & 0.765 & 0.499 & 0.622 & 0.549 & 0.576\\
\hline
\end{tabular}
\label{tab:SupByQuarter}
\end{center}
\end{table}

Somewhat surprisingly, Quarter 16 does not show any decrease in the classification rate compared with Quarters 2-4 (with the exception of Quarter 3, as discussed above). Because Quarter 16 is 4 years after the data the classifiers were trained on, a noticeable decrease would not have been unexpected. However this was not found, and the classifier actually does slightly better on Q4 than on Q2 and Q4. This indicates that the reliability of the combined classifier does not experience significant degradation over even a multiple year long timescale.

\subsection{Attribute of Significance}
\label{sec:AoS}
As was discussed earlier, a final classification is produced based on which classification receives the plurality of picks from the 12 classifiers. As a result, a new attribute, called the "attribute of significance" can be defined as the percent of the classifiers picking the plurality class. This number can be expressed as a percentage ranging from 0-1, and while it should be stressed that it is not the likelihood that the predicted class is correct, it can be related to that likelihood. Generally, a higher attribute of significance should correlate with a greater chance of a correct classification. Conversely, a low attribute of significance may indicate an unusual star, or one that is not any of the pre-supplied classes, explaining why the different classifiers had difficulty coming to a common classification.

For quarter 1, the average star had 44\% of the classifiers picking the eventual classification. Stars that were classified correctly (per the Blomme classifications) had 46.4\% of the classifiers picking correctly, while incorrectly classified stars had only 39.3\% classifiers picking the eventual classification. This difference seems small, but not insignificant, indicating that the number of classifications making up the majority is a useful indication of correct classification.

Table \ref{tab:CutOffCheck} shows another way of visualizing this analysis. As the number of classifiers in the majority drops, the classifications rate hovers around 80-90\% until the number of classifiers drops below six out of the 12. Then the rate at which stars are correctly classifies plummets dramatically. Thus, one could consider those stars with five or fewer classifiers picking the class to be "uncertain" and those with six or more to be "certain." Given these definitions, the ensemble classifier matches the Blomme classifications on an impressive 82\% of the stars it is certain of, which comprise 77\% of the sample. Slightly different choices of the border between certain and uncertain produce similar results. 
\begin{table}
\begin{center}
\caption[Quarter 1 Classification Rate By Attribute of Signifiance]{Quarter 1 Classification rate and attribute of significance. Columns are the absolute number of classifiers producing the majority opinion, their relative percentage of the total, the raw number of correct classifications with the corresponding number of classifiers, the raw total number of stars with the corresponding number of classifiers, and the resulting classification rate respectively.}
\begin{tabular}{|ccccc|}
\hline
Num & Percent & Correct & Total & Rate\\
\hline
12 & 1.000 & 645 & 746 & 0.865\\
11 & 0.917 & 3785 & 4310 & 0.878\\
10 & 0.833 & 8459 & 9254 & 0.914\\
9 & 0.750 & 13759 & 14954 & 0.920\\
8 & 0.667 & 18106 & 20063 & 0.902\\
7 & 0.583 & 21436 & 25932 & 0.827\\
6 & 0.500 & 21573 & 31160 & 0.692\\
5 & 0.417 & 13724 & 22701 & 0.605\\
4 & 0.333 & 4425 & 8564 & 0.517\\
3 & 0.250 & 176 & 965 & 0.182\\
2 & 0.167 & 0 & 21 & 0.000\\
1 & 0.083 & 0 & 0 & n/a\\
\hline
\end{tabular}
\label{tab:CutOffCheck}
\end{center}
\end{table}

Table \ref{tab:CutOffCheckQ2} shows the same analysis for the Quarter 2 data. Quarter 2's general poorer classification rates are clearly visible, but the same pattern of correlation between Attribute of Significance and classification rate, as well as a sudden drop-off when less than half of the classifiers pick the same classification is apparent. Using the same definition of "certainty" from before, only 36\% of the stars are "certain," and on those the classifier matches the Blomme classification 67\% of time. This is worse than Q1, but still considerably higher than the classification rate on the entire sample. 

\begin{table}
\begin{center}
\caption[Quarter 2 Classification Rate By Attribute of Significance]{Quarter 2 Classification rate and Attribute of Significance. Columns are the absolute number of classifiers producing the majority opinion, their relative percentage of the total, the raw number of correct classifications with the corresponding number of classifiers, the raw total number of stars with the corresponding number of classifiers, and the resulting classification rate respectively.}
\begin{tabular}{|ccccc|}
\hline
Num & Percent & Correct & Total & Rate\\
\hline
12 & 1.000 & 0 & 0 & n/a\\
11 & 0.917 & 11 & 12 & 0.917\\
10 & 0.833 & 274 & 349 & 0.785\\
9 & 0.750 & 1291 & 1735 & 0.744\\
8 & 0.667 & 3642 & 4885 & 0.746\\
7 & 0.583 & 9304 & 13003 & 0.716\\
6 & 0.500 & 18427 & 29432 & 0.626\\
5 & 0.417 & 23297 & 47363 & 0.492\\
4 & 0.333 & 11818 & 36328 & 0.325\\
3 & 0.250 & 1228 & 5668 & 0.217\\
2 & 0.167 & 21 & 138 & 0.152\\
1 & 0.083 & 0 & 0 & n/a \\
\hline
\end{tabular}
\label{tab:CutOffCheckQ2}
\end{center}
\end{table}

Overall, this new "Attribute of Significance" is quite useful. It can produce a list of stars that are quite likely to be correctly classified. This ability to distinguish between stars with good or poor classifications is likely to be quite useful to future researchers using any results produced.  

In contrast, those stars with very low numbers of classifiers predicting the majority may also be of interest for further study for different reasons. Perhaps these stars are not typical members of any of the known classes of stars, but are still variables (non-MISC class), and thus of interest as outliers, hybrids, or previously unknown classes. 

One final possibility is using this method as another means of analyzing the efficacy of the ensemble classifier, independent of outside data, such as either the Blomme or the other classifications. Looking at the tables above, the attribute of significance seems to show a normal-like distribution, centred around different numbers of classifiers. One method of rating the classifier could then be where that central point is. For example, the Q1 results, with their mean number of classifiers around 6.95 is better than Q2, which has a mean of 5.22. This indicates that the Q1 results are more reliable. 

While training data are still needed to train the classifier, this method provides a test independent of that training data on the reliability of the resulting classifier, something that may be of great interest in future work in this field and with newer telescopes. 

\subsection{Outside Classifications}
The Q1 classifications were also compared with the outside manual classification sample (described at the end of section \ref{sec:TrainingData}), and a 66\% match was found, which is actually higher than the 54.6\% classification rate for Blomme's original data. This seems to imply that our combined classifier is actually learning correct patterns, even with possibly poor training data. It is important to remember, however, that the manual classifications are not a random sample, and have significant biases in the numbers of different types of stars. Thus, a direct comparison does not necessarily prove that one classifier is superior to another. Figure \ref{fig:MeMiscComparison} shows the confusion matrix.

\begin{figure}
\includegraphics[width=0.99 \columnwidth]{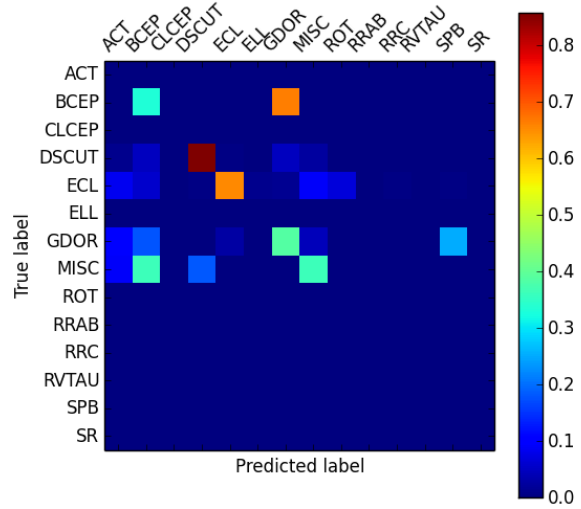}
\caption[Confusion Matrix of Ensemble Classifier vs Manual Classifications]{Heat map of the scaled confusion matrix from the ensemble classifier and the collection of manual classifications found in the literature. The colour displays the percent of all objects with a given true classification that were assigned the listed predicted classification. The literature sample is significantly biased towards ECL stars. The ensemble classifier overall correctly classifies 66\% of the stars, a more than 10 percentage point improvement over the classifications in Blomme et al. The classifier does particularly well on DSCUT, while having difficulty distinguishing GDOR and BCEP}
\label{fig:MeMiscComparison}
\end{figure}

\section{Conclusions and Future Work}
\label{sec:Conclusions}
Classifications for each of the roughly 150,000 stars observed by {\em Kepler} were produced, separating the stars into one of 14 variable star classes. These classifications enable future research into specific types of stars to be better able to focus their studies. The details of how the classifications are produced also enhance our understanding of the differences between these different classes.

We have shown the value of ensemble methods of characterization and classification. In particular, we used a combination of three methods of characterizing stars, Fourier Transformation, SAX, and statistical and photometric data with three very different supervised learning algorithms, Random Forests, Bayes Networks, and Neural Networks. Previous works on Kepler and other similar projects have primarily used just the Fourier Transform and Statistical and Photometric data, and have relied on a single supervised learning algorithm, typically Random Forests and/or Decision Trees. 

This approach has been shown to have significant value. Compared with the classifications by \citet{Blomme_2011}, the only other attempt at classifying all of the Kepler stars, the ensemble method matched their classifications roughly 50-70\% of the time. At the same time, when comparing both classifications with a limited collection of independent, third-party classifications found in the literature, the ensemble methodology showed a 10 percentage point increase in classification accuracy, from roughly 55\% to 65\%.

The classification also provided several interesting results. The Random Forest analysis provided a test on attribute significance that was useful for unsupervised learning. The Bayesian Network provided a confirmation of some of these results, as well as a new way of visualizing the attribute inter-relationships.

\subsection{Future Work}
The related field of unsupervised learning is also of great interest. While supervised learning is designed to classify new instances of objects into already known classifications, unsupervised learning has goals of finding new classifications, understanding relationships between objects in a sample, and other, more interesting and complicated understanding. Simultaneously to this research, we have performed several areas of unsupervised learning which have produced interesting and promising results that should be published soon.

In addition, there are several branches of future work available. First, finding or producing better training and testing data will improve the classifier's results and further validate the advantages of ensemble characterization and classification. This involves human classification of more Kepler data, as well as training on outside data similar to the works of previous studies. 

Another aspect of classification of great interest is semi-supervised learning. This general type of methods involve classifying a small number of the sample, but leaving the rest unclassified. The algorithm is then allowed to find new or existing classifications. The goal is maintain the flexibility and strengths of unsupervised learning, while gaining strength by using a small number of classifications that may either already exist or can be found accurately through minimal effort, typically through manual classification. 

This method could be ideal for data like that from Kepler and future data sets, where the sheer number of new light-curves is impractical to classify manually. The other methods, either using automatically classified data (as we did here), or training on light curves produced by different telescopes (as done by Blomme), both have weaknesses. Semi-supervised learning allows training on a small number of manually classified stars, potentially avoiding these issues. For a recent survey on the state of research of semi-supervised learning, see \citet{Zhu_2006}.

Similarly, active learning is another area of future research in this area that might greatly enhance classification with only minimal requirement for expert human classification. In active learning, the algorithm queries for specific objects to be manually classified. The specific objects are chosen based to most efficiently determine the class boundaries.

Improving the ensemble classifier and training on multiple sources will also render the classifier more general. This will allow it be used on new astronomical time-series datasets forthcoming in the next generation of telescopes, such as GAIA and LSST. 

\section{Acknowledgments}
The author's would like to thank Dr. Mike Summers, Dr. Joseph Weingartner, and Dr. Ruixin Yang for their help, assistance, and suggestions in developing and refining this work.

We thank the anonymous referee for many valuable comments and suggestions. 

This paper includes data collected by the Kepler mission. Funding for the Kepler mission is provided by the NASA Science Mission directorate.

Some/all of the data presented in this paper were obtained from the Mikulski Archive for Space Telescopes (MAST). STScI is operated by the Association of Universities for Research in Astronomy, Inc., under NASA contract NAS5-26555. Support for MAST for non-HST data is provided by the NASA Office of Space Science via grant NNX09AF08G and by other grants and contracts.

The Bayesian Network images and models described in this paper were created using the GeNIe modeling environment developed by the Decision Systems Laboratory of the University of Pittsburgh and available at http://genie.sis.pitt.edu/.
\bibliography{mn-jour,DissertationRefs}

\section{Additional Figures}
\label{sec:AddFigs} 
 A sample light curve is presented for each category of variable star. These light curves were chosen as representative examples from the Kepler data, and are unprocessed beyond the standard Kepler pipeline. In particular, they have not been de-trended, as was done before any actual analysis, and thus any linear trend should be considered an artifact of the telescope. All of the light curves are from the Quarter 1 (Q1) data. 
 
 The figures below show a plot of the SAP (Single Aperture Photometry) flux versus time. The SAP flux is the calibrated summation of all flux falling on a given pixel, in units of electrons per second. The time axis is barycentric Julian day minus 2454833.0
 
\begin{figure}
\includegraphics[width=0.5\textwidth]{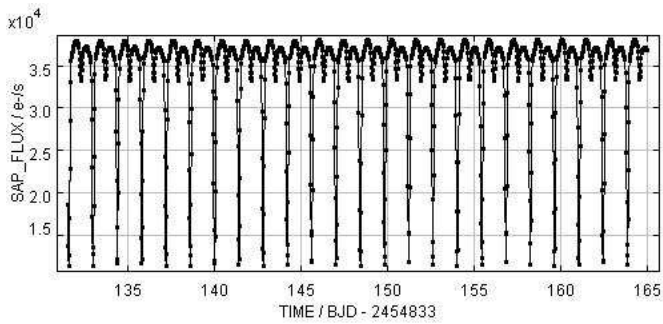}
\caption[Eclipsing Variable Star KID 2305372]{Example Eclipsing Variable Star(ECL) KID 2305372. Note the regular sharp dips, during eclipses. The longer dips are typically when the larger, primary star is eclipsed, and the shorter dips when the secondary star is eclipsed.}
\label{fig:ECLSample}
\end{figure}

\begin{figure}
\includegraphics[width=0.5\textwidth]{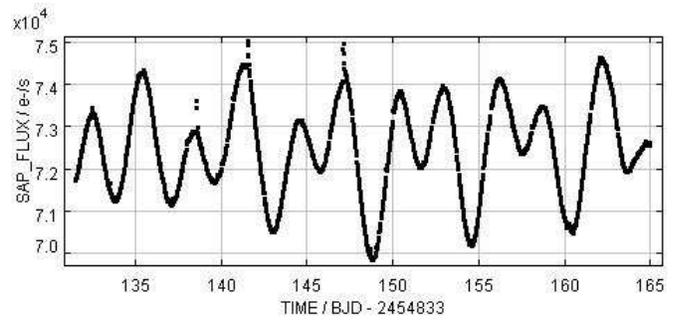}
\caption[Ellipsoidal star KID 3528198.]{Example Ellipsoidal star KID 3528198.}
\label{fig:ELLSample}
\end{figure}

\begin{figure}
\includegraphics[width=0.5\textwidth]{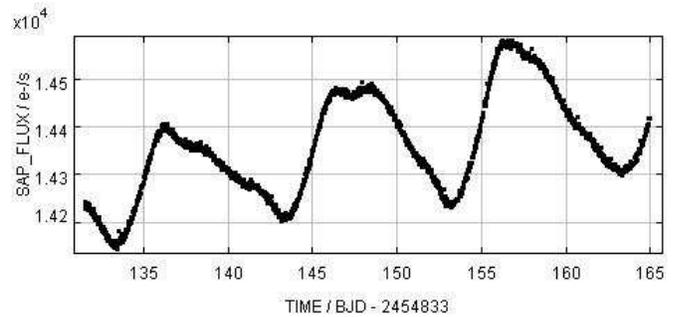}
\caption[Rotational star KID 4066629]{Example Rotational star KID 4066629. Note the clear periodicity over a long time span (the period of this star appears to be about 10 days). The linear trend is an artifact of the telescope and was removed for the actual data analysis.}
\label{fig:ROTSample}
\end{figure}

\begin{figure}
\includegraphics[width=0.5\textwidth]{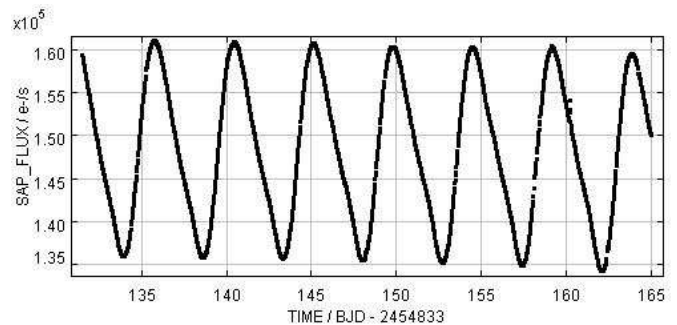}
\caption[Classical Cepheid KID 1573138]{Example Classical Cepheid (CLCEP) star KID 1573138. These stars have very obvious, significant, and regular variability, with a characteristic "shark-fin" curve.}
\label{fig:CLCEPSample}
\end{figure}

\begin{figure}
\includegraphics[width=0.5\textwidth]{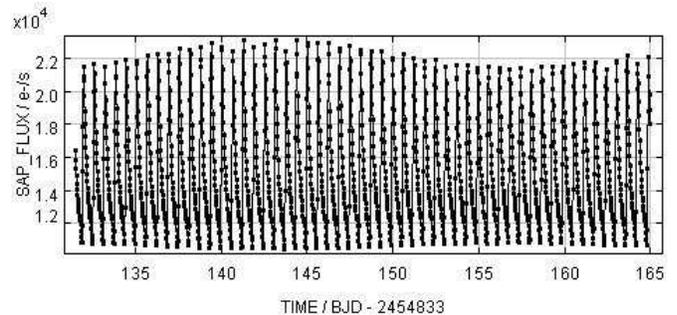}
\caption[RR Lyrae AB KID 5559631]{Example RR Lyrae AB (RRAB) star KID 5559631. }
\label{fig:RRABSample}
\end{figure}

\begin{figure}
\includegraphics[width=0.5\textwidth]{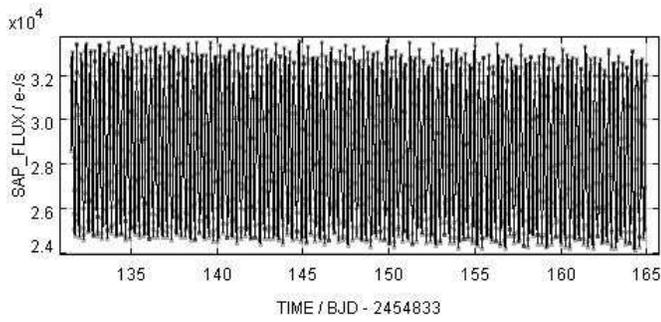}
\caption[RR Lyrae C KID 5520878]{Example RR Lyrae C (RRC) star KID 5520878.}
\label{fig:RRCSample}
\end{figure}

\begin{figure}
\includegraphics[width=0.5\textwidth]{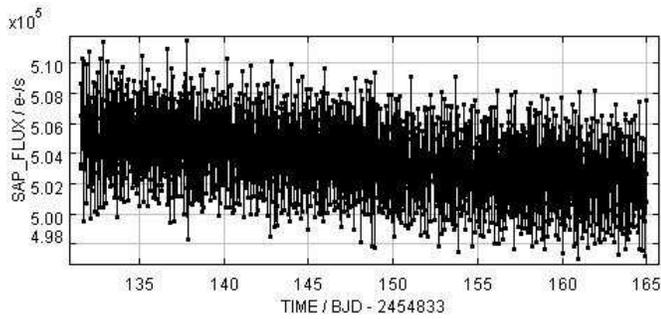}
\caption[$\delta$ Scuti KID 2303365]{Example $\delta$ Scuti (DSCUT) star KID 2303365.}
\label{fig:DSCUTSample}
\end{figure}

\begin{figure}
\includegraphics[width=0.5\textwidth]{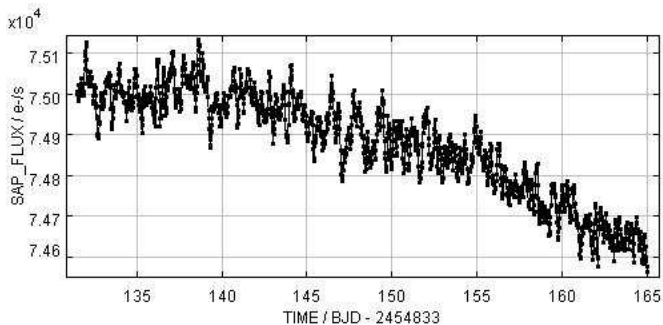}
\caption[$\beta$ Cephei KID 5095336]{Example $\beta$ Cephei (BCEP) star KID 5095336. The rapid, shallow and irregular variability of characteristic of this class of variable star. }
\label{fig:BCEPSample}
\end{figure}

\begin{figure}
\includegraphics[width=0.5\textwidth]{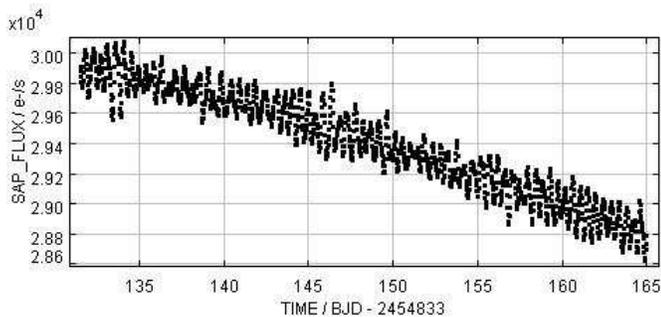}
\caption[$\gamma$ Doradus KID 2449383]{Example $\gamma$ Doradus (GDOR) star KID 2449383. As with the other stars, the sloping linear trend is likely an artifact and was removed during detrending. GDORs have low magnitude variability due to non-radial gravitational wave oscillations.}
\label{fig:GDORSample}
\end{figure}

\begin{figure}
\includegraphics[width=0.5\textwidth]{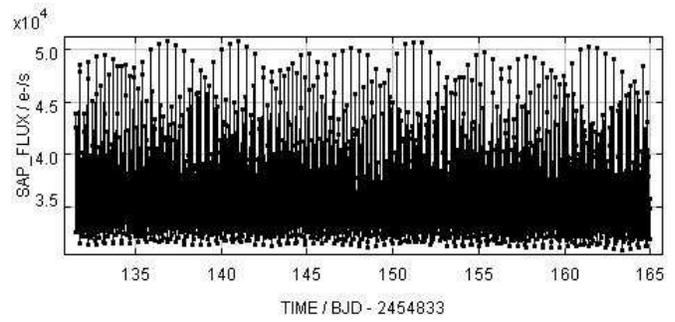}
\caption[RV Tauri KID 5950759]{Example RV Tauri (RVTAU) star KID 5950759. The characteristic regular changing maxima is quite clear.}
\label{fig:RVTAUSample}
\end{figure}

\begin{figure}
\includegraphics[width=0.5\textwidth]{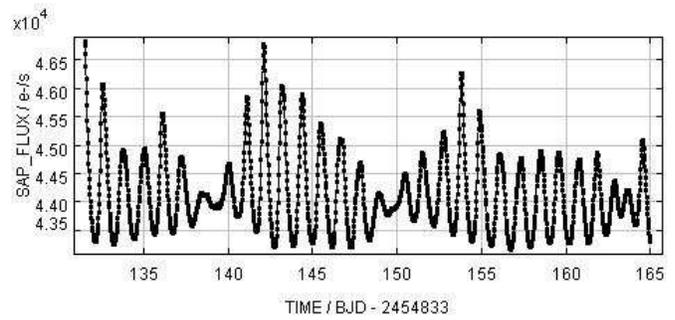}
\caption[Slowly Pulsating B Star KID 3654076]{Example Slowly Pulsating B (SPB) star KID 3654076. The period is quite clear and regular, with the magnitude of pulsating also changing steadily.}
\label{fig:SPBSample}
\end{figure}

\begin{figure}
\includegraphics[width=0.5\textwidth]{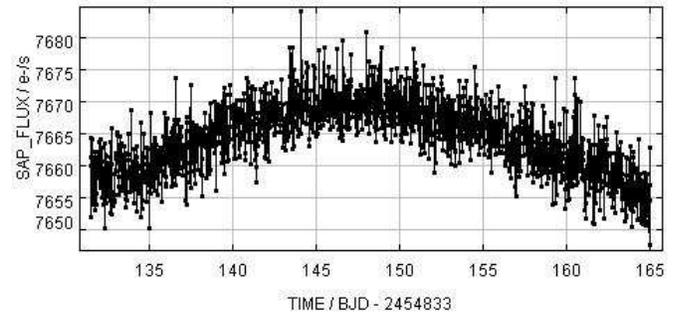}
\caption[Semi Regular Star KID 7451258]{Example Semi-Regular (SR) star KID 7451258. This star appears to be experiencing a long pulsation (Period of around 60 days), along with rapid pulsation of order less than 1 day.}
\label{fig:SRSample}
\end{figure}

\begin{figure}
\includegraphics[width=0.5\textwidth]{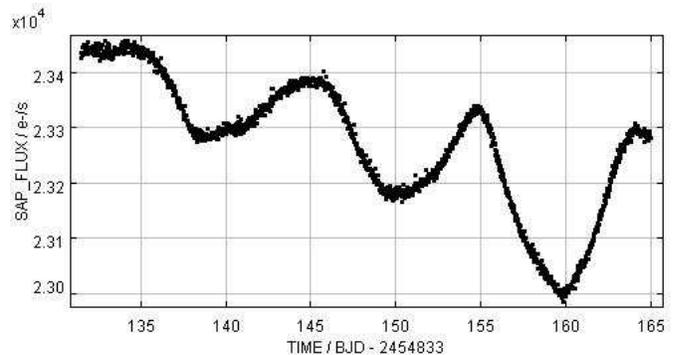}
\caption[Active Star KID 6119964]{Example Activity (ACT) star KID 6119964.}
\label{fig:ACTSample}
\end{figure}

\begin{figure}
\includegraphics[width=0.5\textwidth]{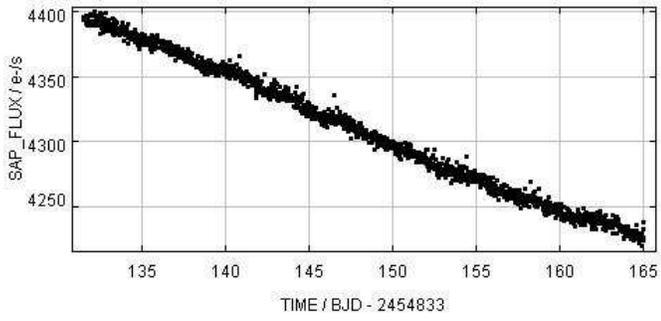}
\caption[Non Variable star KID 1718761]{Non Variable (MISC) star KID 1718761. The light curve is mostly flat, with an artificial linear trend that was removed in later processing.}
\label{fig:MISCSample}
\end{figure}

\end{document}